\documentclass[journal]{IEEEtran}
\ifCLASSINFOpdf

\else

\fi
\usepackage{amsmath}
\usepackage{graphicx}
\usepackage{subfig}
\usepackage{diagbox}
\usepackage{caption}
\usepackage{epstopdf}
\usepackage{cite}
\usepackage{amsfonts}
\usepackage{amssymb}
\usepackage{algorithmic,algorithm}
\usepackage{xcolor}
\begin{document}

\title{A {N}ovel K-Repetition {D}esign for {S}{C}{M}{A}}

\author{Ke Lai, Zilong Liu \IEEEmembership{Senior Member,~IEEE}, Jing Lei, Lei Wen, Gaojie Chen \IEEEmembership{Senior Member,~IEEE} and Pei Xiao \IEEEmembership{Senior Member,~IEEE} 
	\thanks{K. Lai, J. Lei, L. Wen are with Department of Communication Engineering, College of Electronic Science and Engineering, National University of Defence technology. (e-mail: laike12@nudt.edu.cn; leijing@nudt.edu.cn).		 	
    Zilong Liu is with School of Computer Science and Electronics Engineering, University of Essex, UK (e-mail: zilong.liu@essex.ac.uk).    
	G. Chen is with the Department of Engineering, University of Leicester, Leicester LE1 7RH, U.K. (e-mail: Gaojie.Chen@leicester.ac.uk).	
    P. Xiao is with State Key Laboratory of ISN, Xidian University, Xi’an, 710071, China. (email: pxiao2022@163.com).}
}
\maketitle

\begin{abstract}
This work presents a novel K-Repetition based HARQ scheme for LDPC coded uplink SCMA by employing a network coding (NC) principle to encode different packets, where K-Repetition is an emerging technique (recommended in 3GPP Release 15) for enhanced reliability and reduced latency in future massive machine-type communication. Such a scheme is referred to as the NC aided K-repetition SCMA (NCK-SCMA). We introduce a joint iterative detection algorithm for improved detection of the data from the proposed LDPC coded NCK-SCMA systems. Simulation results demonstrate the benefits of NCK-SCMA with higher throughput and improved reliability over the conventional K-Repetition SCMA.
\end{abstract}


\section{Introduction}
\IEEEPARstart  {T}he trend is that future wireless communication systems are rapidly evolving towards providing communications over massive number of user equipments (UEs), which is referred to as machine-type communications (MTC)\cite{Guo20216G}. An emerging paradigm for MTC is non-orthogonal multiple access (NOMA) which enables higher spectral efficiency by overloading multiple users upon finite resource nodes. In this work, we focus on sparse code multiple access (SCMA), which is a promising code-domain NOMA scheme for the next generation MTC systems\cite{Yu2021Sparse}. In order to correct as many data errors as possible, whilst at the same time, attain faster communication speed, a standard technique is Hybrid Automatic Repeat reQuest (HARQ) \cite{Hybrid2021Ahmed}. In 3GPP Release 15, K-Repetition has been introduced as an advanced HARQ scheme\cite{3GPP2020}, by sending repeated packets at the base station (BS) for enlarged diversity and reduced latency. 


Although numerous works concerning enhanced error probability of SCMA have been studied in recent years \cite{Liu2021Sparse,Long2016A}, improving the reliability with HARQ has rarely been investigated. In \cite{Long2016A}, a blanking based HARQ scheme for SCMA is proposed, while in \cite{Zeina2020On}, the performance of HARQ with blankingis carefully analyzed for both power-domain NOMA and SCMA\footnote{HARQ with blanking enforces the UEs without detection errors to be silent, i.e., transmitting no information, in the retransmissions, leading to reduced spectral efficiency.}. However, these two works are tailored for the convetional HARQ, such as HARQ with chase combining (HARQ-CC) and HARQ with incremental redundancy (HARQ-IR) based NOMA. For K-Repetition, system level simulations capturing the main performance influencing factors are reported in \cite{Jacobsen2019System}. An analytical expression of the success probability of K-Repetition is derived in \cite{Liu2021Analyzing} from stochastic geometry perspective. Moreover, as reported in \cite{Choi2021Krep}, random network coding (NC) has been employed in the K-Repetition scheme to generate distinguished linear combinations of preambles so as to enhance the transmission reliability. 

Different from \cite{Choi2021Krep} that improves error rate performance of K-Repetition scheme through resolving pramble collision using NC, this work is dedicated to improving the data transmission and detection for a novel K-Repetition based SCMA (K-SCMA) system over LDPC coded uplink channels, which is referred to as NC aided K-repetition SCMA (NCK-SCMA). We aim to go one-step further by applying NC to encode multiple transmitted LDPC coded packets/blocks, before SCMA encoding is conducted. By doing so, we show that larger amount of coding gain can be attained. Furthermore, to recover data from LDPC coded NCK-SCMA with HARQ retransmissions, a joint iterative detector based on message passing algorithm (MPA) is also proposed. To fully exploit the coding gain of NC and diversity of K-Repetition, a major novelty of this work stems from our proposed soft combination of packets, yielding  different message passing strategies should be made regarding for the states of previous transmissions. Extensive simulation results show the significant improvement of throughput and packet error rate (PER) of NCK-SCMA compared to K-SCMA.

 \emph{Notations}: The operation $[\cdot]^T$ denote transpose of a matrix. $\oplus$ represents the XOR operation. $\lceil\cdot\rceil$ and $\vert \cdot \vert$ are the ceil function and 1-norm respectively. $a\backslash b$ indicates all the elements in set $a$ except $b$.
\section{System model}
In this section, K-Repetition based system model of the proposed scheme is presented in detail.
\subsection{K-Repetition SCMA}
\begin{figure}[!t]
	\centerline{\includegraphics[height=1.4in, width=3.6in]{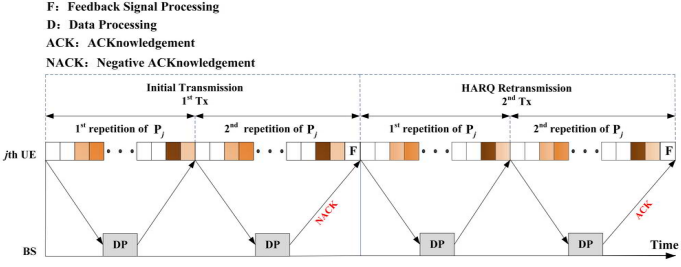}}
	\caption{Illustration of K-SCMA with $K = 2$.}
	\label{fig_K-SCMA}
\end{figure}

Let us consider a single-cell network where each uplink device transmits data to their serving BS using SCMA. Assume that  $J$ users transmit over $R$ allocated resource elements ($J>R$). An SCMA codeword for the $j$th user can be represented as a $R$-dimensional complex vector $\mathbf{X}_j = [X_{j,1}, \cdots, X_{j,R}]^{T}$, where $X_{j,r} \in \mathbb{C}$, $r \in \left\{1, 2, \cdots, R\right\}$,  $j \in \left\{1, 2, \cdots, J\right\}$, $ \mathbb{C}$ denotes the complex field, and $\mathbf{X}_j$ is selected from a pre-defined codebook $\mathcal{X}_j$ with cardinality of $\vert\mathcal{X}_j\vert = M = 2^b$, where $b$ bits map to an SCMA codewor. In K-SCMA, $K$ consecutive replicas of the same packet are transmitted without waiting for the feedback. Fig. \ref{fig_K-SCMA} presents an example of K-SCMA when $K=2$. It can be seen from the figure, the instantaneous input message of the $j$th UE is $\mathbf{P}_j$ which consists of $N$ bits after LDPC coding and bit-level interleaving. $K = 2$ replicas are prepared to be carried out in consecutive transmission time intervals (TTIs) for SCMA encoding. Hence, after SCMA mapping, each packet contains $L =  \lceil \frac{N}{b} \rceil$ SCMA codewords. At the receiver, data processing (DP) is performed by using MPA and soft combinations after receiving $K$ repetitions. Retransmissions will be conducted if a NACK is received from HARQ feedback (F). 


\subsection{Proposed NC based K-Repetition SCMA (NCK-SCMA)}
\begin{figure}[!t]
	\centerline{\includegraphics[height=2.5in, width=3.2in]{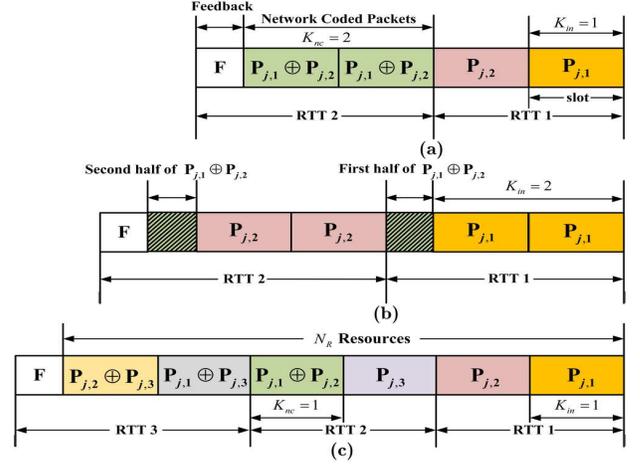}}
	\caption{Illustration of NCK-SCMA with $K_{eq} = 3$. (a) Type-A with $T = 2, K_{in} = 1$; (b) Type-B with $T = 2, K_{in} = 2$; (c) Type-C with $T = 3, K_{in} = 1$.}
	\label{fig_NCK-SCMA}
\end{figure}

Different from K-SCMA which simply transmits $K$ replicas, NCK-SCMA leverages NC to combine $T$ transmitted packets based on XOR operation to harness the coding gain. Let us assume that $T$ different packets are to be sent in several time slots. To ease understanding, examples of three different types of NCK-SCMA are presented in Fig. \ref{fig_NCK-SCMA}. To proceed, let us define $K_{in}$ and $K_{nc}$ as the numbers of initial repetitions and network coded (NCed) repetitions, respectively. As for Type-A NCK-SCMA in Fig. \ref{fig_NCK-SCMA}(a), after transmiting $T = 2$ packets $\mathbf{P}_{j,1}$ and $\mathbf{P}_{j,2}$ with $K_{in} = 1$ in the first round-trip time (RTT), $K_{nc} = 2$ replicas of NCed packets, i.e., $\mathbf{P}_{j,1} \oplus \mathbf{P}_{j,2}$ together with the feedback signal are transmitted in the next RTT. In this case, it is equivalent to send both $\mathbf{P}_{j,1}$ and $\mathbf{P}_{j,2}$ three times, and thus the equivalent value of $K$, denoted by $K_{eq}$, is 3, i.e., $K_{eq}=3$.  
Fig. \ref{fig_NCK-SCMA}(b) demonstrates the Type-B NCK-SCMA that obtains the same $K_{eq}$ through introducing a slightly higher signaling overhead as the NCed packets are treated as redundancy. It is shown that one $\mathbf{P}_{j,1} \oplus \mathbf{P}_{j,2}$ packet is separated into two different parts which are sent with $K_{in} = 2$ repetitions in each RTT. 
To elaborate further, Fig. \ref{fig_NCK-SCMA}(c) plots Type-C NCK-SCMA. As can be observed from the figure, $T = 3$ packets are to transmit. Let us denote by $W$ the total number of NCed combinations of the $T$ packets. Since XOR operations are performed in pairs, we have $W=C(T,2)$, where $C(\cdot,\cdot)$ denotes the binomial coefficient. In Type-C NCK-SCMA shown in Fig. 2(c), we have $W=3$. As such, $K_{eq} = 3$ can be achieved with $K_{in}=K_{nc}=1$. 
Let $N_R$ be the number of resources (TTIs) finishing a complete NCK-SCMA or K-SCMA with the same $T$. One can see that, Type-A and Type-C NCK-SCMA use the same $N_R$ as K-SCMA with $ K = 2$ to transmit $T = 2$ and $T = 3$ different packets, respectively, with only one feedback. For instance, in Type-C NCK-SCMA, $N_R = 6$, whereas the same $N_R$ should be also utilized to transmit $T=3$ packets in K-SCMA with $K=2$.

Based on the above discussions, we consider a generalized version of NCK-SCMA.
For a NCK-SCMA that intends to transmit $T$ packets, after generating $K_{in}$ initial repetitions of each packet $\mathbf{P}_{j,t}$, where $t\in\{1,2,\cdots,T\}$, NC will be performed to combine $T$ packets in pairs. Assume further that $t_{\alpha}, t_{\beta} \in \left\{1,2,\cdots, T\right\}$ are the indices of two different packets before performing NC, and let $\Xi$ be the set of all the possible $\{\alpha,\beta\}$, such that the NCed packets can be represented as $\mathbf{P}_{j,t_{\alpha}} \oplus \mathbf{P}_{j,t_{\beta}}$, and $\vert \Xi \vert = W$. Since NC is employed, it is equivalent to have $K_{eq}$ repetitions, where $K_{eq}$ is
\begin{equation}
	K_{eq} = K_{in} + \frac{2\left(N_R - TK_{in}\right)}{T}.  \label{K_{eq}}
\end{equation}
 It should be noted that K-SCMA is a special case of NCK-SCMA by defining $K_{in}=K$, $T = 1$ and $K_{nc}=0$, and thus one can show that $K \leq K_{eq}$ by substituting $KT$ and $K$ into $N_R$ and $K_{in}$ in  \eqref{K_{eq}}, respectively. Furthermore, according to \eqref{K_{eq}}, $K_{nc}$ should satisfy $ K_{nc} = \frac{K_{eq} - K_{in}}{T -1}$ under the constraint of $TK_{in} + WK_{nc}= N_R$. From above discussions, it is clear that we can use $K_{eq}$, $T$, and $K_{in}$ to uniquely characterize a NCK-SCMA. For simlicity, $(K_{eq}, T, K_{in})$ is used to represent a NCK-SCMA with a specific SCMA signature matrix.
The advantage of NCK-SCMA can be intuitively interpreted from above discussions, i.e., a higher $K_{eq}$ can be achieved compared to K-SCMA under the condition that approximately the same $N_R$ are utilized. This allows NCK-SCMA to further exploit the coding gain to enhance throughput and reliability. 

\section{Proposed Detection for LDPC coded NCK-SCMA}
In this section, we develop a joint detection algorithm for LDPC coded NCK-SCMA. In our proposed joint iterative detector, belief messages generated in the current and previous transmissions are updated iteratively among the LDPC, NC, and SCMA decoders.
Therefore, the proposed scheme can fully utilize the advantages of NC and HARQ retransmission compared to K-SCMA. For clarity, we define SCMA and LDPC variable nodes as SVN $sv_{l,j}$ and LVN $lv_{k,n}$, where $l\in\left\{1, 2, \cdots,N_RL\right\}$, $k\in\left\{1, 2, \cdots,JT\right\}$, and $n\in\left\{1, 2, \cdots,N\right\}$ denote the indices of SCMA codewords, LDPC detectors, and LVNs, respectively. Furthermore, a virtual node for network decoding and message conversion is denoted as NC check node (NCN) $nc_{\eta}^{t,j}$, where $\eta\in\left\{1, 2,\cdots,N\right\}$ is the index of NCN for the $t$th transmitted packet of the $j$th UE. In addition, the function nodes (FN) of SCMA and parity check nodes (PN) of LDPC update should be also taken into consideration. Before elaborating on the update rules of these nodes, we define the messages as follows.
\begin{itemize}
	\item  $I_{{l}, {r\to j}}$: Message passed from FN $f_{l,r}$ to SVN $sv_{l,j}$;
\end{itemize}
\begin{itemize}
	\item  $G_{{l}, {j\to r}}$: Message passed from SVN $sv_{l,j}$ to FN $f_{l,r}$;
\end{itemize}
\begin{itemize}
	\item  $Q_{{k}, {c\to n}}$: Message passed from PN $p_{k,c}$ to LVN $lv_{k,n}$;
\end{itemize}
\begin{itemize}
	\item  $S_{{k}, {n\to c}}$: Message passed from LVN $lv_{k,n}$ to PN $p_{k,c}$;
\end{itemize}
\begin{itemize}
	\item  $\Lambda^{t,j}_{\eta\to \hat{j}}$, $\Lambda^{t,j}_{\eta\to {n}}$: Message passed from NCN $nc_{\eta}^{t,j}$ to SVN $sv_{l,j}$, and NCN $nc_{\eta}^{t,j}$ to LVN $lv_{j,n}$,
\end{itemize}
where $c\in\left\{1,2,\cdots,C\right\}$,  $C$ is the number of PN for LDPC code, and $\hat{j} \in \left\{1,2,\cdots,\cdots,JN_RL\right\}$ is the index of SVN for each packet. 
\subsubsection{Initialization}
In this paper, the messages of SCMA and LDPC are defined in the probability and logorithm domain, respectively. Thereby, the initial messages are $I_{{l}, {r\to j}} = 1/M$, $Q_{{k}, {c\to n}} = 0$, $\Lambda^{t,j}_{\eta \to\hat{j}} = 1/M$, and  $\Lambda^{t,j}_{\eta \to n} = 0$.
\subsubsection{FN Update and PN Update}
After receiving the superimposed signals of NCK-SCMA at FNs, the detection will start from FN and PN simultaneously.
In our proposed algorithm, FN and PN only exchange information with SVN and LVN, respectively; hence, the update of FN and PN are the same as the conventional SCMA and LDPC, which can be written as:
\begin{equation}
	I_{{l}, {r\to u}}\left(\mathbf{X}_j\right) =\frac{1}{\pi N_0}e^{\left(-d^{j,r}_l/N_0\right)} \cdot \prod\limits_{j\in \xi_r \backslash u}G_{{l}, {j\to r}}\left(\mathbf{X}_j\right), \label{FN_update} 
\end{equation}
where $d^{j,r}_l = \frac{1}{N_0/2} \vert Y_{l,r}-\sum_{r \in \xi_r}h^{j,r}_lX_{j,r} \vert^2$, $h^{j,r}_l$ denotes the channel coefficient between the $j$th UE and the $r$th resource of the $l$th codeword, $Y_{l,r}$ the received signal at the $r$th FN of the $l$th codeword, $N_0$ the noise variance, and $\xi_r$ the node set consisting of SVNs connect to the $r$th FN. Meanwhile, the PN update is given by:
\begin{equation}
		Q_{{k,c\to {i}}} = 2 \times \tanh^{-1}\Bigg(\prod_{n\in \phi_c \backslash i}\tanh(S_{{k,n}\to {c}})\Bigg), \label{CN_update}
\end{equation}
 where  $\phi_c$ is the set that contains all the LVNs connecting to the $c$th PN. The messages in both \eqref{FN_update} and \eqref{CN_update} should be normalized in each iteration for the purpose of numerical stability.
\subsubsection{SVN Update}
In the update of SVN, messages that come from FN and NCN should be involved. For clarity, we first consider SVN update for the non-NCed packets, and thus
\begin{equation}
	G_{l_{1}+ \lceil\frac{\eta}{b}\rceil, {j\to v}}\left(\mathbf{X}_j\right) = \prod\limits_{r\in \zeta_j\backslash v} I_{l_{1}+ \lceil\frac{\eta}{b}\rceil, {r\to j}}\left(\mathbf{X}_j\right) \cdot \mathcal{M}^{-1}\left(\Lambda^{t,j}_ {\eta\to\tilde{j} } \right), \label{SVN_update1}
\end{equation}
where $l_{1} = (t-1)K_{in}L + (k_{in}-1)L , k_{in}\in \left\{1,2,\cdots, K_{in}\right\}$, $\tilde{j} = J(l_1 + \lceil\frac{\eta}{b}\rceil -1) + j$, $\zeta_j$ is the FN set of nodes that connect to the $j$th SVN, and $\mathcal{M}^{-1}$ is the inverse marginalization which transform bit-wise soft information to symbol level\footnote{For ceil function, note that there exist $b$ different elements that satify $\lceil\frac{n_1}{b}\rceil = \cdots = \lceil\frac{n_b}{b}\rceil$ for a certain $n\in\mathbb{N}^{+}$, which indicates that $b$ different messages can be generated from the same node indexed by $\lceil\frac{n}{b}\rceil$ in this paper.} in this paper. However, since NC is employed in NCK-SCMA, to improve the performance of iterative detection, the soft information comes from NCN should be re-encoded to generate messages for the NCed packets. Hence, 
\begin{equation}
\begin{aligned}
	G_{l_{2}+ \lceil\frac{\eta}{b}\rceil, {j\to v}}\left(\mathbf{X}_j\right) = \prod_{r\in \zeta_j\backslash v} &I_{l_{2}+ \lceil\frac{\eta}{b}\rceil, {r\to j}}\left(\mathbf{X}_j\right) \\ &\cdot \mathcal{M}^{-1}\left(\Lambda^{t_{\alpha},j}_ {\eta\to \tilde{j}_{\alpha}} \boxplus \Lambda^{t_{\beta},j}_{\eta\to \tilde{j}_{\beta}} \right), \label{SVN_update2}
	\end{aligned}
\end{equation}
 where $l_{2} = TLK_{in} + (w-1)K_{nc}L + (k_{nc}-1)L , k_{nc}\in \left\{1,\cdots, K_{nc}\right\}$, and $\tilde{j}_{\Xi} = J((t_{\Xi}-1)LK_{in} + \lceil\frac{\eta}{b}\rceil - 1) + j$.
Moreover, $\boxplus$ is the soft XOR operator that is defined as \cite{Salamat2020Short}: 
\begin{equation}
	\begin{aligned}
		L_1 \boxplus L_2 &= 2\tanh^{-1}\left(\tanh\left(L_1/2\right) \cdot \tanh\left(L_2/2\right)\right).\label{Soft_XOR}
	\end{aligned}
\end{equation}
\subsubsection{LVN Update}
Likewise, the update of LVN involves the information coming from PN and NCN. As such, for a certain packet $t$, the update rule of LVN can be represented as 
\begin{equation}
	S_{t+(j-1)T, n \to q} = \sum\limits_{c\in \psi_{n}\backslash q} Q_{t+(j-1)T, c \to n} + \pi_{t,j}^{-1}\left(\Lambda^{t, j}_{\eta  \to n}\right), 
\end{equation}
where $\psi_{n}$ is the node set whose elements connect to the $n$th LVN, and $\pi_{t,j}^{-1}$ denotes the bit level de-interleaver operation.
\begin{figure}[!t]
	\centerline{\includegraphics[height=1.8in, width=3.5in]{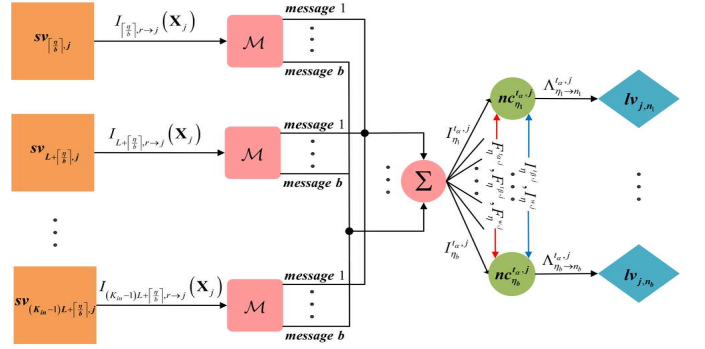}}
	\caption{Partial view of NCN update for a non-NCed packet indexed by $t_{\alpha} = 1$.}
	\label{fig_NCN_up}
\end{figure}
\subsubsection{NCN Update}
It is clear that NCN bridges the SCMA detection and LDPC decoding, and thus the update of NCN should be carefully designed to fully utilize the NC packets. A partial view of the NCN update is presented in Fig. \ref{fig_NCN_up}. As can be seen from the left side of the figure, to update $\Lambda^{t, j}_{{\eta \to {n}}}$, we should first consider the messages at SVN after soft combination of the $K_{in}$ initial repetitions\footnote{It is worth noting that the subscript of SVN in Fig. \ref{fig_NCN_up} is obtained by substituting $t_{\alpha} = 1$ into $l_1$.}, which can be expressed as:
\begin{equation}
	I_\eta^{t,j} = \sum_{k_{in} = 1}^{K_{in}} \Bigg[\mathcal{M}\Bigg(\prod\limits_{r\in \zeta_j } I_{\lceil\frac{\eta}{b}\rceil + l_{1}, {r\to j}} \left(\mathbf{X}_j\right)\Bigg)\Bigg],\label{NCN_update1}
\end{equation}
where $\mathcal{M}$ is the marginalization that converts $M$-dimensional message vector to $b$ messages. To process NCN update, the $n$th message of NCed packets after $K_{nc}$ times soft combination is also necessary, which can be represented as:  
\begin{equation}
	I_\eta^{w,j} = \sum_{k_{nc} = 1}^{K_{nc}} \Bigg[\mathcal{M}\Bigg(\prod\limits_{r\in \zeta_j } I_{\lceil\frac{\eta}{b}\rceil + l_{2}, {r\to j}}\left(\mathbf{X}_j\right)\Bigg)\Bigg].\label{NCN_update2}
\end{equation} 
It is also shown in Fig. \ref{fig_NCN_up} that messages $I_{\eta_1}^{t_{\alpha},j}$ to $I_{\eta_b}^{t_{\alpha},j}$ pass to $b$ differernt NCNs after marginalization and combining $K_{in}$ soft information of packet $t_{\alpha}$. Furthermore, once NCK-SCMA retransmission is requested, the soft information generating from previous unsuccessful detections can be utilized. It is assumed that $F_\eta^{w,j}$, $F_\eta^{t_{\alpha},j}$, and $F_\eta^{t_{\beta},j}$ are the previous soft information of NCed packet $w$ and its corresponding packets $t_{\alpha}$ and $t_{\beta}$. After receiving all the soft information at NCNs, the receiver can output $\Lambda^{t_{\alpha}, j}_{\eta \to n}$.
In order to properly use these messages, the following three cases should be taken into consideration. For simplicity, we take the message $\Lambda^{t_{\alpha}, j}_{\eta \to n}$ as an example to demonstrate the NCN update, and the updating rules of $\Lambda^{t_{\beta}, j}_{{\eta\to {n}}}$ can be derived similarly.

\underline{Case 1:} \emph{The previous $T$ packets all fail.} In this case, as long as the maximum transmission rounds $N_{re}$ has not been reached, the retransmission is necessary. To ensure the accuracy of soft information, we have
\begin{equation}
	\Lambda^{t_{\alpha}, j}_{\eta \to n} = \big[(I_\eta^{t_{\beta},j} + F_\eta^{t_{\beta},j}) \boxplus (I_\eta^{w,j} + F_\eta^{w,j})\big] + I_\eta^{t_{\alpha},j} +  F_\eta^{t_{\alpha},j}. \label{case1}
\end{equation} 
The updating rule in Case 1 follows from the maximum-ratio combining (MRC) principle, and thus the messages generated in previous and new transmission can be fully utilized.

\underline{Case 2:} \emph{The previous $T$ packets all successful.} In this instance, $T$ new packets are sent at the current RTT, which indicates that $ F_\eta^{t_{\beta},j} = F_\eta^{w,j} = F_\eta^{t_{\alpha},j} = 0$, for $\forall t,w,j,\eta$ in \eqref{case1}. Therefore, the NCN update can be written as $\Lambda^{t_{\alpha}, j}_{\eta \to {n}} = (I_\eta^{t_{\beta},j} \boxplus I_\eta^{w,j} )+ I_\eta^{t_{\alpha},j}$.

\underline{Case 3:} \emph{The previous $T$ packets partially fail.} In this case, the messages  corresponding to newly transmitted packets update follow the rule in Case 2, while the NCN update for those packets that are unsuccessfully detected is:
\begin{equation}
		\Lambda^{t_{\alpha}, j}_{{\eta\to {n}}} = \underbrace{I_\eta^{t_{\alpha},j} +  F_\eta^{t_{\alpha},j}}_{\text{I}} + \underbrace{\big[I_\eta^{t_{\beta},j} \boxplus I_\eta^{w,j}\big] +  \big[F_\eta^{t_{\beta},j} \boxplus F_\eta^{w,j}\big]}_{\text{II}}.\label{case3}
\end{equation} 
As can be observed from part I of Eq. \eqref{case3}, to avoid distortion of soft information caused by soft XOR in \eqref{Soft_XOR}, we combine the messages of packets that suffer from failure in previous and current transmissions at first. After that, as can be seen from Part II, the information comes from NCed packets involved in the NCN update to further enhance the reliability of this case. 

On the other hand, the messages passing from NCN to SVN is
$\Lambda^{t, j}_{\eta\to \hat{j}} = \pi_{t,j}\left(Q_{t+(j-1)T,n}\right)$, 
where $Q_{k,n} = \sum\limits_{c\in \psi_{n}} Q_{k,c\to n}$, $\pi_{t,j}$ is the interleaver for the $t$th packet of UE $j$.

\subsubsection{Decision and Output Messages}
In the proposed joint detection scheme, the final decision is made at LVN, and the accumulated logorithm likelihood ratio (LLR) is:
\begin{equation}
	Q_{t+(j-1)T,n} = \sum\limits_{c\in \psi_{n}} Q_{t+(j-1)T,c\to n} + \Lambda^{t, j}_{\eta \to n}, \label{NCN_update} 
\end{equation}
whereas the LLRs for the possible HARQ retransmission output at SVN, are given by $\eqref{NCN_update1}$ and $\eqref{NCN_update2}$. The iteration stops once the syndromes are zeros or reaching to the maximum number of iteration.
\section{Numerical results and analysis}
In this section, the throughput and PER performances of K-SCMA and NCK-SCMA are evaluated to demonstrate the effectiveness of the proposed scheme. Two different SCMA signature matrices and codebooks having orders of  ``$4\times6$" and ``$5\times10$" are considered \cite{Liu2021Sparse}. Moreover, two 5G new ratio (NR) LDPC codes, as specified in \cite{3GPP2018}, with rates of 1/2 and 5/6, respectively, are used for simulation. We consider Rayleigh fading channel and the maximum iteration number of 50.
As indicated in \cite{Long2016A,Nadeem2021Nonorthogonal}, the average throughput is treated as an important criterion to evaluate the performance of the proposed NCK-SCMA. Based on the definition in \cite{Nadeem2021Nonorthogonal}, we have
 \begin{equation}
 	\theta = \dfrac{T_{correct}}{T_{total}},  \label{throughput}
 \end{equation}
where $T_{correct}$ and $T_{total}$ are the number of correctly decoded and the total number of transmitted packets, respectively. In addition, to make fair comparisons with K-SCMA, we also define the average utilized resources transmitting the same number of packets as $\bar{N}_R = N_R/T$. Specifically, for K-SCMA, we have $K = N_R = \bar{N}_R$. 
 \begin{figure}[!h]
 	\centerline{\includegraphics[height=2.1in, width=3.0in]{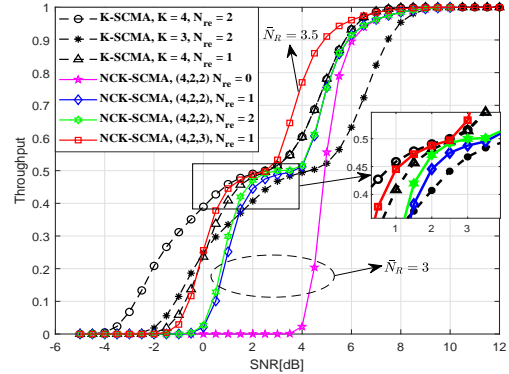}}
	\caption{Throughput comparison for ``$4 \times 6$" K-SCMA and NCK-SCMA with code rate of 5/6 , and $N = 260$.}
	\label{fig_Throughput_Comp_4x6_R_5_6}
\end{figure}

Fig. \ref{fig_Throughput_Comp_4x6_R_5_6} compares the throughput of ``$4 \times 6$" NCK-SCMA and K-SCMA with code rate of 5/6. As can be observed from the figure, Type-A NCK-SCMA with (4,2,2) and $\bar{N}_R=3$ can achieve higher throughput than K-SCMA with $K=3$ without any retransmissions when SNR $\geq5$ dB. By performing retransmission using NCK-SCMA with (4,2,2), the throughput can be significantly improved at low and medium SNRs. Nevertheless, K-SCMA with $K = 3$ and $N_{re}=2$ still outperforms NCK-SCMA with (4,2,2) in the relatively low SNR region. It is worth noting that the increase of $N_{re}$ leads to marginal improvement of throughput in NCK-SCMA as can be seen from the figure. This can be attributed to the fact that the soft XOR operation may have negative impacts on the effectiveness of soft information, especially at low SNRs. As a result, it may lead to error propagation in the proposed joint detection algorithm when reusing the soft information in previous transmissions. It should be noted that Type-A NCK-SCMA with $K_{eq}=4$ can approach the throughput of K-SCMA with $K=4$ with fewer resource consumptions in the medium-to-high SNR regime. Moreover, Type-B NCK-SCMA with (4,2,3) and $\bar{N}_R=3.5$ is also simulated, in which a redundant NCed packet is added. Therefore, higher throughput can be achieved compared to K-SCMA with $K=4$.
 \begin{figure}[!t]
	\centerline{\includegraphics[height=2.1in, width=3.0in]{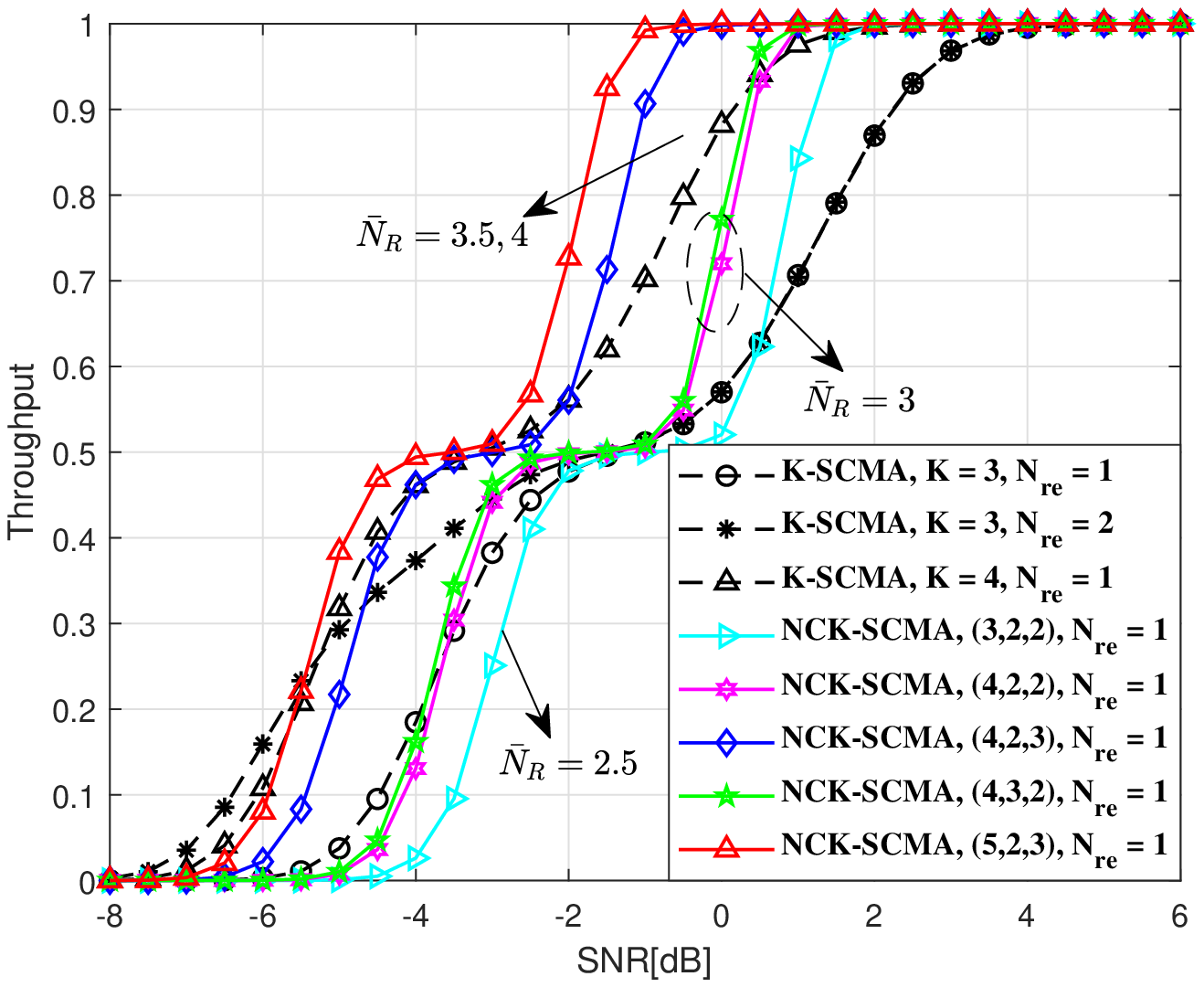}}
	\caption{Throughput comparison for ``$5\times10$" K-SCMA and NCK-SCMA with code rate of 1/2, and $N = 270$.}
	\label{fig_Throughput_Comp_5x10_R_1_2}
\end{figure}

Fig. \ref{fig_Throughput_Comp_5x10_R_1_2} investigates the throughput performances of  ``$5\times10$" K-SCMA and NCK-SCMA with code rate of 1/2. We can observe that the Type-C NCK-SCMA with (4,3,2) slightly performs Type-A NCK-SCMA with (4,2,2) under the same $N_{re}=1$ and $\bar{N}_R=3$. This follows from the fact that higher coding gain may be achieved by Type-B NCK-SCMA as more NCed packets are transmitted. Moreover, both of them show superiority over the K-SCMA with $K=3$ and $N_{re}=1$. When SNR $\geq-3$ dB, NCK-SCMA with (4,2,2) also outperforms albeit with a slight degradation in the region of SNR $<-3$ dB. For NCK-SCMA with (4,2,3) and (5,2,3) that possesses $\bar{N}_R = 3.5$ and $\bar{N}_R=4$, respectively, we can also observe that though slight throughput loss can be seen in the low SNR regime, both of them exhibit far higher throughput compared to K-SCMA with $K=4$ at medium-to-high SNRs. That being said, the advantages of NCK-SCMA in throughput become more significant with the increase of SNR.

As for PER performances, if a single bit of a packet cannot be correctly decoded after $N_{re}$ HARQ retransmissions, the packet is deemed to be in error. Based on this definition, we evaluate the PER performances of K-SCMA and NCK-SCMA in this subsection.
 \begin{figure}[!t]
	\centerline{\includegraphics[height=2.2in, width=3.5in]{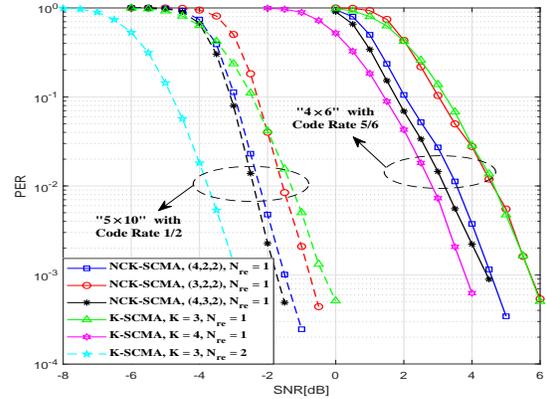}}
	\caption{PER comparison for K-SCMA and NCK-SCMA.}
	\label{fig_PER_Comp_M4}
\end{figure}
Fig. \ref{fig_PER_Comp_M4} shows the PER performances of K-SCMA and NCK-SCMA under two different LDPC codes and overloading factors. It can be seen from the figure, for both $``4 \times6"$ and  $``5 \times10"$ NCK-SCMA with (4,3,2), improved PER performance is achieved across all SNRs compared to that with (4,2,2), which is consistent with the throughput results discussed above. Furthermore, we can observe that NCK-SCMA with (3,2,2) has approximately the same PER performance as K-SCMA with $K = 3$ at code rate of $5/6$, while NCK-SCMA with (3,2,2) can achieve 0.3 dB gain at PER =$10^{-3}$. We can also observe that the gap between K-SCMA with $K=4$ and NCK-SCMA with (4,3,2) is 0.3 dB at PER = $10^{-2}$. However, at code rate of $1/2$, as $N_{re}$ increases, the gap between K-SCMA and NCK-SCMA becomes more significant. 
\section{Conclusion}
In this paper, we have introduced a novel NC aided K-repetition scheme for SCMA called NCK-SCMA. According to the approaches obtaining the same $K_{eq}$, we have presented three different types of NCK-SCMA. To recover data from such a network-LDPC coded SCMA with repetitions and retransmissions, we have further developed a joint SCMA-NC-LDPC detector. Our results demonstrated that NCK-SCMA achieves higher throughput at medium-to-high SNR regime with fewer resources and retransmissions, while an enhanced PER can be also obtained under the same $\bar{N}_R$ and $N_{re}$.

\ifCLASSOPTIONcaptionsoff
\newpage
\fi
\bibliographystyle{IEEEtran}
\bibliography{Krep_for_SCMA}

\end{document}